\documentclass[]{rsos}

\usepackage[flushleft]{threeparttable}
\usepackage{enumitem}
\usepackage{graphicx}
\usepackage{subfig}
\usepackage{amsmath,caption,threeparttable,booktabs}
\usepackage[tableposition=top]{caption}
\usepackage{booktabs}
\usepackage{bm}

\begin{document}

\title{Cultural Investment and Urban Socio-Economic Development: A Geo-Social Network Approach}

\author{
Xiao Zhou$^{1,2}$, Desislava Hristova$^{2}$, Anastasios Noulas$^{3}$, Cecilia Mascolo$^{2}$ and Max Sklar$^{4}$}

\address{$^{1}$Dept. of Land Economy, University of Cambridge, UK \\
$^{2}$Computer Laboratory, University of Cambridge, UK \\
$^{3}$Center for Data Science, New York University, USA \\
$^{4}$Foursquare Labs, New York, USA}

\subject{Computer science}


\corres{Xiao Zhou\\
\email{xz331@cam.ac.uk}}

\begin{abstract}
Being able to assess the impact of government-led investment onto socio-economic indicators in cities has long been an important target of urban planning. However, due to the lack of large-scale data with a fine spatio-temporal resolution, there have been limitations in terms of how planners can track the impact and measure the effectiveness of cultural investment in small urban areas. Taking advantage of nearly 4 million transition records for three years in London from a popular location-based social network service, Foursquare, we study how the socio-economic impact of government cultural expenditure can be detected and predicted. Our analysis shows that network indicators such as average clustering coefficient or centrality can be exploited to estimate the likelihood of local growth in response to cultural investment. We subsequently integrate these features in supervised learning models to infer socio-economic deprivation changes for London's neighbourhoods. This research presents how geo-social and mobile services can be used as a proxy to track and predict socio-economic deprivation changes as government financial effort is put in developing urban areas and thus gives evidence and suggestions for further policy-making and investment optimisation.
\end{abstract}

\maketitle
\section{Introduction}
In 1997, the striking 'Bilbao miracle' created by Guggenheim Museum not only provided Bilbao, a depressed northern Spanish port town, with a dramatic socio-economic growth, but also demonstrated that cities can blossom with cultural investment \cite{1,2}. Even though the ability of culture to promote local regeneration has received general acceptance, large-scale evaluation and prediction of its impact are still not widely practised. The potential of network science in offering insight on deprivation dynamics \cite{3} along with the millions of human mobility traces made available by location-based applications has so far been largely untapped in culture-led regeneration studies. In this paper, we propose a new fusion of techniques using geo-social network data from Foursquare\footnote{https://foursquare.com} to quantify the effect of cultural investment on the urban regeneration process and predict its outcome in London's neighbourhoods. 

Culture-led urban regeneration, as one of the main branches of urban regeneration, has received increasing attention globally in recent decades and been applied by a number of governments as a boost to revitalising depressed urban areas. Historically, it was in the 1970s when culture was first used as a catalyst to accelerate urban regeneration and by the late 1980s when the term 'culture-led regeneration' started to emerge in literature \cite{2}. Ever since then, the significant role that culture can play in urban regeneration has been widely discussed by researchers. In \cite{4}, Keddie  pointed out culture's effectiveness in reducing the deprivation level and promoting 'social mixing' for urban areas; Vickery stated that culture can be utilized by cities to improve existing environment, attract tourism, increase employment, and reinforce civic pride \cite{2}. In addition to the benefits mentioned above, another 'by-product' of culture-led regeneration is creating the city branding, which is thought to be particularly attractive to those international metropolises with an expectation to make the city an alluring base so as to promote its functional role in the global economy \cite{5,6}. Realising the positive effects that might be brought, a growing number of cities have begun to put more effort and allocate more financial resources to culture to promote urban regeneration. London, the city we choose to study in this research, is no exception. Despite local government budgets experiencing considerable pressure as a result of the central government funding cuts in recent years, the local authorities in London remain significant supporters of arts and culture. In 2013/14 for example, the spending of London boroughs on arts and culture was \pounds220.5 million, representing around 3 per cent of the total local authority spending in the city, in comparison with 2.2 per cent nationally \cite{7}. 

However, how to measure the socio-economic impact of culture-related policy and expenditure is still an open question. Conventionally, the investigation of socio-economic deprivation for urban areas has largely relied on government statistics, with data generally obtained through the traditional survey. It is usually costly to implement and takes a few years to carry out each time. With an aim to overcome this limitation, researchers have recently started to mine low-cost, real-time, and fine-grained new data sources for socio-economic deprivation study. For instance, Eagle et al \cite{3} discovered a high correlation between call network diversity and urban area deprivation using call records data; Louail et al \cite{8} showed the application of bank card transaction data on socio-economic inequalities study in cities; Quercia et al \cite{9} found the topic of tweets and the deprivation level of urban areas are correlated; Smith et al \cite{10} used Oyster Card data to identify areas of high deprivation level in London; Quercia and Saez\cite{11} explored the relationship between the presence of certain Foursquare venues with social deprivation; Venerandi et al\cite{12} used Foursquare and OpenStreetMap datasets to explore the correlation between urban elements and deprivation of neighbourhoods; And Hristova et al \cite{13} discussed the relationship between the prosperity of people and urban places, and distinguished between different categories and urban geographies using Foursquare and Twitter data. In line with this stream of research, we take advantage of the geo-social network data from Foursquare and show its success in capturing and predicting the socio-economic change related to government cultural expenditure which has vital implications for culture-led urban regeneration in neighbourhoods. 

More specifically, we utilise the spatial network of Foursuare venues formed by the trajectories of users to track the changes of urban areas. The reason why such kind of geo-social network data is used is because cities are complex systems where the effects of regeneration often grow from the bottom up. Different from traditional analyses that have often adopted top-down methodologies which largely ignore local features \cite{14},  network science, a bottom-up approach in nature, provides opportunities to link behaviours of individuals together in a spatio-temporal framework and enables significant insights in local physical and social transitions \cite{14,15}. This makes it possible to observe and understand the ever-changing dynamics of culture-led urban regeneration at fine grain as government financial effort is put into developing urban areas. Recently, some researchers have begun to take advantage of this certain type of data to understand cities. In \cite{16}, Karamshuk et al focused on the problem of optimal retail store placement and explored how the popularity of three retail store chains in New York is shaped; Georgiev et al \cite{17} extracted indicators of the spatial positioning of retailers as well as the mobility trends of users to model the economic impact of the Olympic Games on local businesses in London; And Noulas et al \cite{18} investigated the topological properties of the urban place networks created by mobility data across a large set of 100 cities globally and applied supervised learning algorithms to predict new links between venues. To the best of our knowledge, this is the first work in which collective transition data from geo-social network is used in culture-led regeneration studies. The main contributions of our work include:
\begin{itemize}[leftmargin=0.4cm]
 	\item We propose an innovative approach to giving insights on underlying relationships between socio-economic status, cultural investment and geo-social network properties using a fusion of techniques, including network analysis, statistical analysis, and supervised machine learning.
 	\item We demonstrate how datasets from government and geo-social network with different spatial and temporal granularities can be analysed jointly and produce the inference of local socio-economic change at finer temporal grain than official government statistics.
 	\item We define new metrics on cultural investment and cultural features in geo-social networks to measure the priority level of culture for urban areas and show how the differences in these metrics reflect on the network properties of local areas.
 	\item Applying traditional network metrics to the geo-social graph of transitions between venues on Foursquare, we show that areas with high cultural investment and deprivation level experience significant growth in the following years.
 	\item We prove it feasible to adopt geographic, cultural expenditure, and geo-social network features to predict the binary socio-economic deprivation change for small urban areas with high prediction performance. Our evaluation shows the effectiveness of our prediction models with AUC values of up to 0.85. 
 	\item We evaluate the predictive capability for different classifiers and features, with Naive Bayes and random forests being the classification methods that give the best performance. In terms of the prediction features that work best, geo-social network features as a whole are the most powerful predictors.
\end{itemize}

Overall, our findings open new directions for the detection and prediction of socio-economic conditions in urban environment through collective transition behaviours in geo-social networks. The remainder of this work is structured as follows: after first describing the datasets and metrics that are used, we run a preliminary analysis on London boroughs to get a general view and lay a basis for further investigation. We then outline four hypotheses underpinning our analyses grounded on the preliminary analysis and existing literature, which derive from two key concerns: the relationships between urban socio-economic development, cultural investment, and geo-social network features; the feasibility of predicting socio-economic development through cultural expenditure, geo-social network, and geographic features. We then examine the rationality of our hypotheses using ANOVA analysis and a supervised learning classification framework. We conclude with a discussion, including contributions and limitations of our findings.

\section{Dataset}
In this section, we describe datasets used for the study and present their basic properties. In total, there are three major data sources, which are: socio-economic data, cultural expenditure data, and Foursquare data. 

\subsection{Socio-economic Data}
The dataset used to evaluate socio-economic status for neighbourhoods is the English Indices of Deprivation, an official measure of relative deprivation for small areas (Lower Super Output Areas (LSOA\footnote{https://data.london.gov.uk/dataset/statistical-gis-boundary-files-london})) in England calculated by the Department for Communities and Local Government (DCLG). It is organised across seven sub domains (Health Deprivation and Disability; Employment Deprivation; Income Deprivation; Education, Skills and Training Deprivation; Crime; Barriers to Housing and Services; and Living Environment Deprivation), offers deprivation scores for each and produces the Index of Multiple Deprivation (IMD), which reflects the overall deprivation level. This IMD is the index we particularly focus on in the study to assess socio-economic status of London areas. It has been calculated since the 1970s, and is updated every 3-5 years. The latest version of this index is the IMD 2015\footnote{https://www.gov.uk/government/statistics/english-indices-of-deprivation-2015}, which updates the previous version of the IMD 2010\footnote{https://www.gov.uk/government/statistics/english-indices-of-deprivation-2010}. In this paper, we employ these two versions of the IMD to track changes of deprivation levels, and thus to understand the socio-economic change of London areas through a comparative analysis. In the published data, each LSOA  in England is given an IMD score and is ranked from the most deprived to the least deprived, allowing users to be aware of how much more or less deprived an area compared to another. A range of summary measures is also available for users to describe deprivation for higher-level geographies. It is worth noting that the IMD scores are not directly comparable between years. However, it is possible to compare IMD ranking changes for neighbourhoods between 2010 and 2015 versions to get a view of whether an area became relatively worse or better in terms of socio-economic condition during the period, and how large the change was. A more detailed explanation will be given in 3(a).

\subsection{Cultural Expenditure Data}
The cultural expenditure data utilised in this work is the local authority revenue expenditure and financing derived from DCLG\footnote{https://www.gov.uk/government/collections/local-authority-revenue-expenditure-and-financing}. This dataset is based on returns from all 444 local authorities in England, showing how they spend their money for each financial year. It provides information about the local authority revenue spending on various service areas, one of which is 'cultural and related services'. This specific category of cultural expenditure can be further divided into five sub areas: culture and heritage, recreation and sport, open spaces, tourism, and library service. In this study, revenue spending data for financial years 2010/11, 2011/12, and 2012/13 of all these cultural sub areas are used. 

\subsection{Foursquare Data}
Alongside the two official datasets from government introduced above, we also employ user mobility records and venue information of London through a three-year long dataset from Foursquare. This location-based social network dataset contains 'transitions' (successive pairs of check-ins created by users) occurring within London from January 2011 to December 2013. For each transition, venue IDs and timestamps of both origin and destination are recorded. In addition, information of Foursquare venues including the geographic coordinates, category, creation time and the total number of users that have check-in(s) at each venue is also available. In total, there are 3,992,664 transitions generated between 17,804 venues in London during the study period. In Figure 1, we map these Foursquare venues by parent categories with cultural venues coloured. Here, cultural venues are defined and selected as urban places of arts, media, sports, libraries, museums, parks, play, countryside, built heritage, tourism and creative industries, following the line set by the Office of the Deputy Prime Minister in Regeneration through Culture, Sport and Tourism\footnote{http://www.communities.gov.uk/archived/publications/localgovernment/regenerationthroughculture}. As we can see from the figure, cultural venues tend to be situated in Inner London than outer suburbs. It is also noticeable that the density of Foursquare cultural venues in general is higher for West London than the east of the city.  

\begin{figure}[!h]
	\centering\includegraphics[width=4.8in]{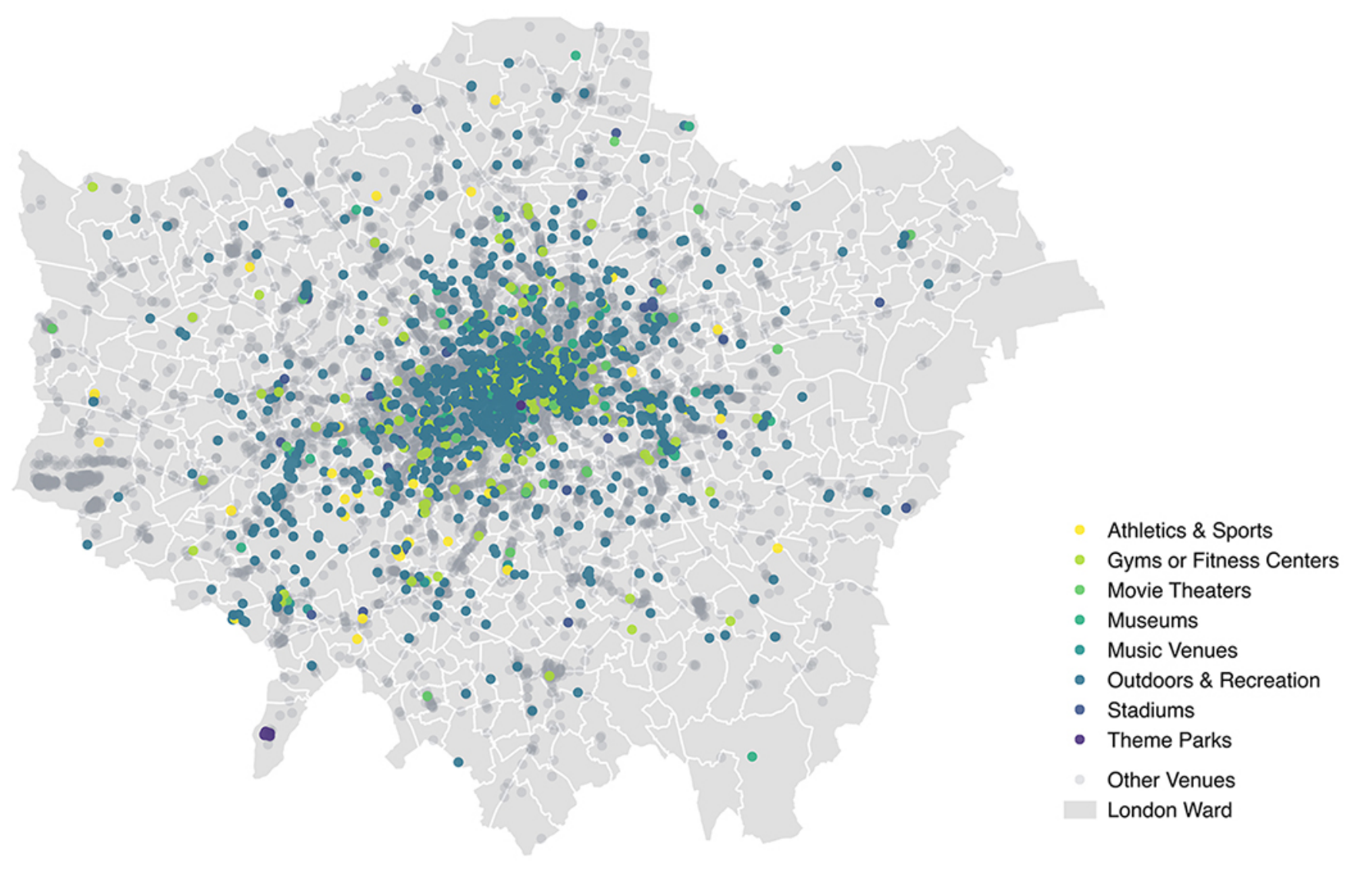}
	\caption{Spatial Distribution of Foursquare Venues in London}
	\label{fig1}
\end{figure}

The Foursquare dataset can be represented as a spatial network of venues connected by transition flows of users for a certain period of time. It is a directed graph where nodes represent start and end venues, while edges correspond to transitions. In the graph, two nodes are linked if at least one trip exists between two venues during the time period. The weight of an edge is proportional to the number of transitions made by all users between the two venues. Since the timestamps of each transition is obtainable, we can study how links are created and how network graph features are changing over time. As one would expect, investments need time to attain an observable effect. The result of cultural expenditure, no matter on the organisation of a music festival, the renovation of a library, or the construction of a new art gallery may take days, months or years to be observable. Here, we assume the impact of cultural expenditure from local authorities can be observed after 9 months on average through Foursquare. Based on this assumption, expenditure and geo-social network data are compared according to time scales in Table 1, where we look at annual snapshots of the data for different years. Formally, we define our yearly dataset as a directed graph $ G_{t}=\left(V_{t},E_{t}\right) $ for $ t=1,2,3 $, which indicate three snapshots in time of the dataset. The set of nodes $ V_{t}=\left\{v_{1},v_{2},v_{3},...,v_{N_{t}}\right\} $ is composed of $ N_{t} $ Foursquare venues and the set of edges $ E_{t} \subseteq V_{t}\times V_{t} $ is composed of pairs of venues that have at least one transition generated between each other during time period $ t $. An edge $ \left(v_{ot},v_{dt}\right) \in E_{t} $ is called a transition edge between $ v_{ot} $ and $ v_{dt} $, where $ v_{ot} $ represents the origin of the transition and $ v_{dt} $ represents the destination. The network properties for each year are shown in Table 2.

\begin{table}
	\centering
	\caption{Comparison Table of Time Scales for Datasets}
	\label{table1}
	\begin{threeparttable}
		\scriptsize
		\normalsize
		\begin{tabular}{cc}
			\toprule
			{\textbf {Expenditure dataset} }& {\textbf{GSN dataset}} 
			\\
			\midrule
			Financial year 2010/11 (Apr 2010 - Mar 2011)&Jan 2011 - Dec 2011 \\ 
			Financial year 2011/12 (Apr 2011 - Mar 2012)&Jan 2012 - Dec 2012 \\
			Financial year 2012/13 (Apr 2012 - Mar 2013)&Jan 2013 - Dec 2013 \\ 
			\bottomrule
		\end{tabular}
	\end{threeparttable}
\end{table}

\begin{table}
	\centering
	\caption{Network Properties at Each Snapshot}
	\label{table2}
	\begin{threeparttable}
		\scriptsize
		\normalsize
		\begin{tabular}{cccccc}
			\toprule
			{\textbf t }& {\textbf{Duration}}  & {$\bm{|V|}$} & {$\bm{|E|}$} & {$\bm{\langle C \rangle}$} & {$\bm{\langle k \rangle}$}
			\\
            \midrule
	     	1 &January 2011 - December 2011 &15832&469229&0.221&59 \\
		    2 &January 2012 - December 2012 &16189&715113&0.228&70 \\
		    3 &January 2013 - December 2013 &17684&742017&0.240&84 \\
		    \bottomrule
	    \end{tabular}
        \begin{tablenotes}
        	\scriptsize
        	\item Number of Nodes $ |V| $, Number of Edges $ |E| $, Average Clustering Coefficient $ \langle C \rangle $, and Average Degree  $ \langle k \rangle $.
	    \end{tablenotes}
    \end{threeparttable}
\end{table}

\subsection{Spatial Unit of Data}
Since the first two datasets are provided by government at different geographic levels initially, here, we introduce the spatial units used in this research and demonstrate how they are applied in our further investigation. This clarification is necessary since understanding how geo-referenced data is aggregated spatially is key to linking the datasets used in the paper. 

The two geographic levels of London areas used in this paper are borough and ward\footnote{https://data.london.gov.uk/dataset/statistical-gis-boundary-files-london}.  
In the research, we perform our exploratory analysis on the aggregate borough level to have a first look and then improve on geographic granularity by using wards as smaller localities for our statistical evaluation and prediction. Here, the reason why ward is selected as the geographic unit for the prediction is because it provides a sufficient training set size for supervised learning models, which is unreachable for the borough (there are totally 32 boroughs and 625 wards in London). As mentioned above, the IMD data is available at several geographic levels, including both borough and ward, by consulting its official document. However, the DCLG only provides expenditure data initially at borough level. We obtain the data for wards by dividing the cultural expenditure of a borough by the number of wards it includes, assuming resources are spent proportionally. As for Foursquare data, check-ins and venues are described at the level of latitude and longitude coordinates. This fine grained spatial representation of activity at the Foursquare data layer allows for standard geographic aggregation methods to be applied in order to attain representations at the ward and borough levels, making therefore the linkage of the three datasets possible.

\section{Metrics}
Leveraging on the described data, we introduce a number of metrics highlighting the advantage that certain neighbourhoods have in terms of cultural expenditure and the properties of geo-social networks for local areas. We also compute a number of geographic features based on the neighbourhood's location and Foursquare venue information. 

\subsection{Geographic Features}
In Figure 2, we map the IMD change for London wards from 2010 to 2015 with sub region information provided. Here, we firstly rank the wards according to their IMD scores from the most deprived to the least in 2010 and 2015, respectively. Then, we subtract the rank in 2010 from that in 2015 for each ward to look at the change, based on which, wards are coloured blue or red, according to whether they became more deprived or less deprived from 2010 to 2015. The larger the change, the darker the shade. From the result, we can see that the ward that experienced the largest improvement locates in East London with an increase of 212 in the IMD ranking. On the other hand, the rank of a ward in Central London dropped most significantly by a number of 190. We can also observe from Figure 2 that areas with similar IMD ranking changes tend to be spatially clustered. And neighbourhoods showing larger improvements in terms of the overall deprivation level are more likely to be in East London. These findings suggest that geographic factors can have an influence on the deprivation level change of urban areas. In this case, we involve a number of geographic features as metrics, including the sub region a ward belongs to, its area size, and how far it is from the city centre. 

\begin{figure}[!h]
	\centering\includegraphics[width=5in]{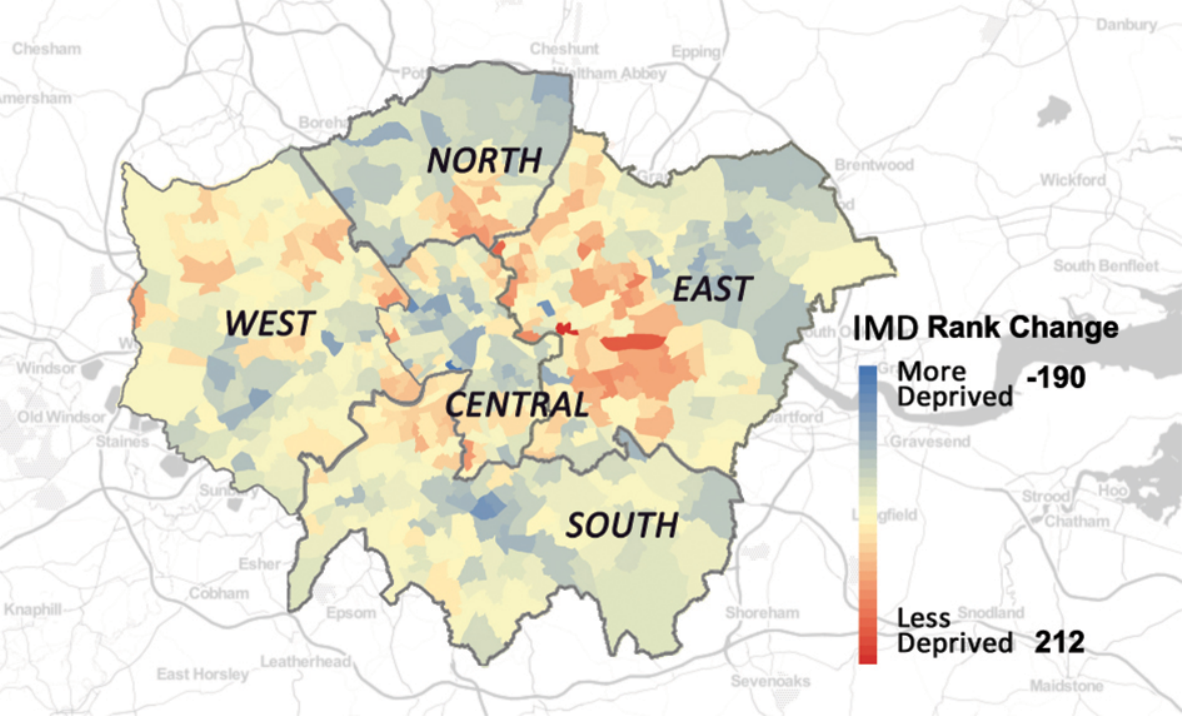}
	\caption{Map of IMD Change for London Wards and Sub Regions}
	\label{fig2}
\end{figure}

In addition, some metrics about Foursquare venues with their geographic properties considered are also given, including the number of venues created (\textit{VC}) and venue created density (\textit{VCD}). Different from the concept of node we described in the previous section, \textit{VC} is defined as the total number of venues emerging in an area during a certain period of time and estimated on the basis of creation time information obtained from Foursquare venue profiles. To divide the \textit{VC} by \textit{area size}, we get \textit{VCD}, which represents the average number of new venues created in an urban area per square kilometre. 

\subsection{Network Metrics}
The network measures applied in this research are in-degree centrality (\textit{IC}), out-degree centrality (\textit{OC}), in-degree/out-degree ratio (\textit{IOR}), and average clustering coefficient (\textit{ACC}). In-degree centrality of an area $ i $ represents how many in-flow transitions the nodes of area $ i $ receive from nodes of other areas. In contrast, out-degree centrality measures how many out-flow transitions start from $ i $, but flow to other areas. We also introduce a metric called \textit{IOR}, which indicates the ratio of in-flow transitions over out-flow transitions. If the \textit{IOR} of an area is high, it means the area is more likely to be an attractive place to visit for people from other places. For area $ i $, the \textit{IOR} can be calculated by:

\begin{equation}\label{eq1}
IOR_{i} = \frac{IC_{i}}{OC_{i}}
\end{equation}
The local clustering coefficient captures the degree to which the neighbours of a given node are connected with each other. For a node $ i $ with degree $ k_{i} $,  the local clustering coefficient \cite{19} is defined as: 

\begin{equation}\label{eq2}
C_{i} = \frac{L_{i}}{k_{i}(k_{i}-1)}
\end{equation}
where $ L_{i} $ represents the number of edges between the $ k_{i} $ neighbours of node $ i $. Then, the average clustering coefficient, which reflects the overall level of clustering in an area is measured as the average of the local clustering coefficients of all the nodes in it.

\subsection{Growth Rate}
We also introduce growth rate metrics for some features to present changes of urban areas. Take growth rate of nodes ($GRN$) as an example, we define this metric to reveal the temporal change of nodes in geo-social network graphs. If the number of nodes we observe in a network snapshot during a year period $ (t-1) $ is $ N_{t-1} $, and a number of $ N_{t} $ in the subsequent time period $ t $, we calculate the $ GRN $ of graph for $ t $ as:

\begin{equation}\label{eq3}
GRN_{t} = \frac{N_{t}}{N_{t-1}}
\end{equation}

In a similar way, other growth rates measures listed in Table 3 can be obtained. 

\subsection{Cultural Advantage Metrics}

In order to measure how the cultural level of a neighbourhood is higher or lower than the average city, we introduce two cultural advantage metrics which rely instead on the concept of location quotients in economic geography. Location quotients capture regional industry specifics by comparing an area's business composition to that of a larger geographic context (i.e., state or nation) and can be calculated by the following formula:

\begin{equation}\label{eq4}
LQ_{i}^{j} = \frac{q_{i}^{j}}{\sum\nolimits_{j\in J}{q_{i}^{j}}}\cdot\left(\frac{\sum\nolimits_{i\in I}{q_{i}^{j}}}{\sum\nolimits_{i\in I}\sum\nolimits_{j\in J}{q_{i}^{j}}}\right)^{-1}
\end{equation}
In the equation, $LQ_{i}^{j}$ represents location quotients for each industry $ j $ and for each region $ i $; $q_{i}^{j}$ denotes the gross output of industry $j$ in region $i$; $I$ and $J$ are the sets of regions and industries, respectively. 
Here, the values of location quotients vary by region due to its industry makeup and can be interpreted through comparing with 1. If the value is greater than 1, it signalises that the concentration of a certain industry in a particular region is higher than average level and a value less than 1 indicates the industry is relatively scarce in that region \cite{20,21}.

\subsubsection{Cultural Expenditure Advantage}

Inspired by industry location quotient, we define a metric called cultural expenditure advantage ($CEA$) to evaluate the priority of cultural expenditure for a neighbourhood in the city. This metric reflects the extent to which a local authority spends more on culture than the city average level. The $CEA$ for area $i$ in the city can be represented as: 

\begin{equation}\label{eq5}
CEA_{i}=\frac{CE_{i}}{TE_{i}}\cdot\left(\frac{\sum\nolimits_{i\in I}{CE_{i}}}{\sum\nolimits_{i\in I}{TE_{i}}}\right)^{-1}
\end{equation}
where $ CE_{i} $ is the amount of cultural expenditure of neighbourhood $ i $; $ TE_{i} $ is the amount of total expenditure of $ i $; and $I$ is the set of neighbourhoods in the city. Through comparing $ CEA $ with 1, whether the cultural expenditure of an area is higher than the city average can be evaluated.

\subsubsection{Cultural Venue Advantage}
Similar as the $ CEA $ defined previously, a metric of cultural venue advantage ($CVA$) is given to reflect the extent to which a neighbourhood has more cultural venues than the city average. Here, cultural venues includes 8 major categories of Foursquare culture-related places presented in Figure 1. The $CVA$ for neighbourhood $ i $ can be defined as:

\begin{equation}\label{eq6}
CVA_{i}=\frac{CV_{i}}{TV_{i}}\cdot\left(\frac{\sum\nolimits_{i\in I}{CV_{i}}}{\sum\nolimits_{i\in I}{TV_{i}}}\right)^{-1}
\end{equation}
where $CV_{i}$ is the number of cultural venues in $i$ and $TV_{i}$  is the total number of venues in $i$.

A summary of all the metrics used in following analyses and tests are listed in Table 3 with their categories, descriptions, and where they are applied in this work provided.

\begin{table}
	\centering
	\caption{Description of the Variables Used in the Analyses}
	\label{table3}
	\begin{threeparttable}
		\scriptsize
		\normalsize
		\begin{tabular}{>{\centering}p{0.15\columnwidth}>{\centering}p{0.115\columnwidth}p{0.485\columnwidth}p{0.118\columnwidth}}
			\toprule
			{\textbf{Category}} & {\textbf{Metric}} & \multicolumn{1}{c}{\textbf{Description}} & {\textbf{Application}}
			\\
			\midrule
			Initial IMD & \textit{Initial IMD} & Rank of IMD at the beginning &P [H3]\\\hline
			&\textit{Sub Region} &Sub-region of London where a ward locates&[H3] \\
			&\textit{Area}&Size of a ward ($km^2$)&[H3] \\
			&\textit{Distance}&Distance from the centre of London to spatial centre of a ward ($km$)&[H3] \\
			Geographic &\textit{VC} &Number of venues created in an area&P [H1] [H2] \\
			&\textit{VCD} &Number of venues created in an area per $km^2$&[H1] [H2] \\
			&\textit{CVA} &Extent to which an area provides more cultural venues than city average&P \\
			&\textit{GRVC} &Growth rate of venues created number&[H3] \\\hline
			&\textit{N}&Number of nodes for an area&[H1] [H2] \\
			&\textit{IC}&Number of in-flow transitions an area receives from other areas&P [H1] [H2] \\
			&\textit{OC}&Number of out-flow transitions an area receives from other areas&P [H1] [H2] \\
			&\textit{IOR}&Ratio of number of in-flow transitions over out-flow transitions &[H1] [H2] \\
			Network &\textit{ACC}&Degree to which nodes in a ward tend to clustering together&[H1] [H2] \\ 
			&\textit{GRN}&Growth rate of number of nodes&[H3] \\
			&\textit{GRI}&Growth rate of number of in-flow transitions&[H3]\\
			&\textit{GRO}&Growth rate of number of on-flow transitions&[H3]\\
			&\textit{GRIOR}&Growth rate of ratio of in-flow transitions over out-flow transitions&[H3]\\
			&\textit{GRACC}&Growth rate of average clustering coefficient&[H3]\\\hline
			&\textit{CE}&Expenditure on cultural and related services&P\\
			&\textit{CEA}&Extent to which an area spends more on culture than city average&P [H3] \\
			Cultural &\textit{CEOP}&Expenditure on open spaces per capita&[H3] \\
			Expenditure &\textit{CECH}&Expenditure on culture and heritable per capita&[H3] \\
			&\textit{CELS}&Expenditure on library services per capita&[H3] \\ 
			&\textit{CERS}&Expenditure on recreation and sport per capita&[H3] \\
			&\textit{CET}&Expenditure on tourism per capita&[H3] \\
			\bottomrule
		\end{tabular}
		\begin{tablenotes}
			\scriptsize
			\item P represents preliminary analysis.
		\end{tablenotes}
	\end{threeparttable}
\end{table}

\section{Preliminary Analysis}
The exploration of relationships between the IMD, cultural expenditure, and network graph features is the foundation of our prediction task for the socio-economic deprivation change. Before discussing it in more depth on a finer spatial scale, we run a preliminary analysis on London boroughs in this section to visualise the relationship patterns and provide evidence for further investigation. 

In this part of analysis, we investigate how areas with different deprivation levels spent their money on culture at the start of our observation period, how they adjusted their priorities in the following years, and how their network graphs changed accordingly. Firstly, Figure 3 is created to reveal the initial relationship between the IMD score and cultural advantage metrics ($ CEA $ and $ CVA $) in 2010. In this figure, the colour bar on the right presents the IMD score of London boroughs in 2010, where yellow means more deprived and purple indicates less deprived. The IMD score is represented by the circle size, where the larger the circle, the higher level of deprivation the area. In addition,
we partition the figure at $CEA=1$ and $CVA=1$. Through these two axes across the 1, Figure 3 is split into quadrants, allowing us to see where boroughs with different deprivation levels are centralised. As we can observe from this plot, yellow circles cluster in the middle/lower part, suggesting more deprived boroughs spent relatively less on culture-related services and showed average cultural venue advantage at the beginning of the study. 

\begin{figure}[!h]
	\centering\includegraphics[width=3in]{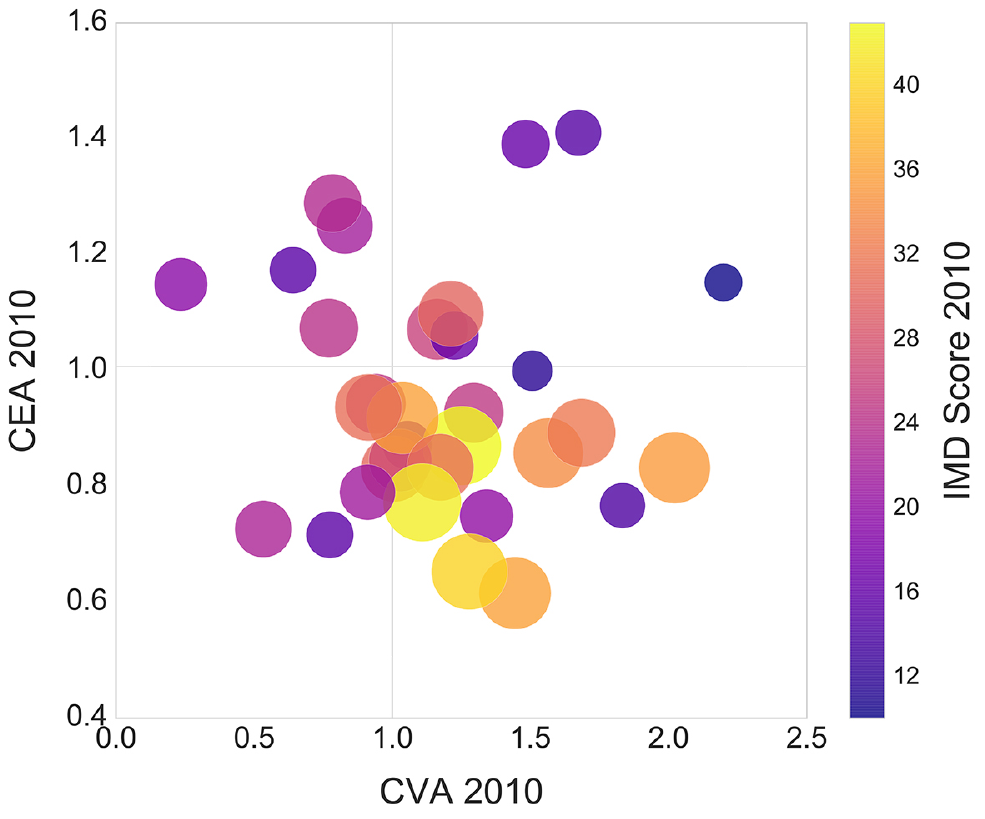}
	\caption{Initial IMD Score, CEA, and CVA of London Borough}
	\label{fig3}
\end{figure}

Then, in Figure 4, we discuss how London boroughs spent their money on culture in the next two years, and how their network and local properties changed. It can be observed that yellow circles, which represent more deprived boroughs, show at the upper right of the charts. In contrast, purple circles, which stand for well-off areas, present in the lower left part of the plots. These signals suggest that compared to prosperous areas, deprived neighbourhoods in London spent more money on culture and had larger number of venues created, and higher in-degree centrality and out-degree centrality showed from 2011 to 2013. Additionally, these trends were more obvious among most and least deprived boroughs compared to average ones. 

\begin{figure}[!h]
	\centering\includegraphics[width=5.3in]{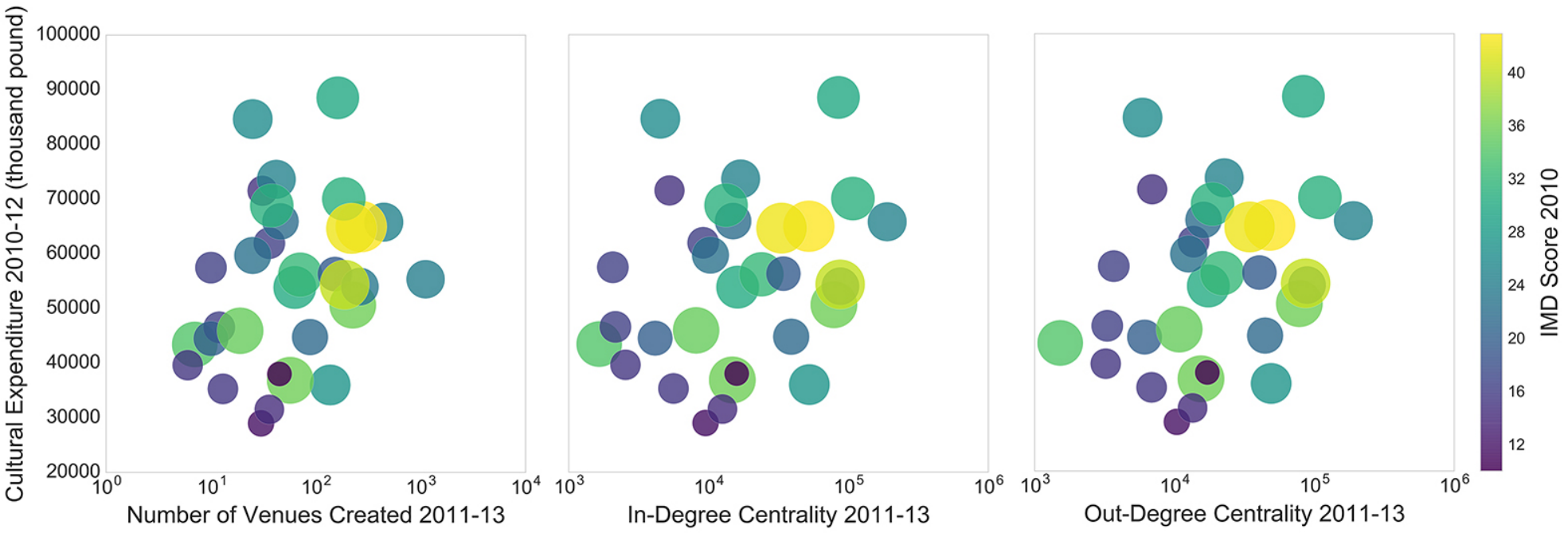}
	\caption{Culture Expenditure, and Foursquare Features Changes of London Boroughs}
	\label{fig4}
\end{figure}

Through the above analysis, we find that initial deprivation level and cultural expenditure strategy may influence the local network graph of urban areas. Furthermore, investing more in cultural and related services seems to have the ability to boost local development for deprived areas. 

In the following sections, the question will be discussed in greater depth by improving on spatial and temporal granularity, and involving statistical and machine learning techniques. To prepare for the analysis, we distinguish London's neighbourhoods on the basis of cultural expenditure priority and the deprivation level. Specifically, wards are firstly grouped into two categories, more deprived and less deprived, according to whether their IMD 2010 deprivation level is higher or lower than the city average. Then, the two groups are further classified according to their cultural spending priorities. If the \textit{CEA} of a ward is more than 1, it is clustered into the more advantaged groups; otherwise, it is put into the less advantaged groups. On the basis of these rules, we can identify four groups of cases outlined in Table 4 and mapped in Figure 5. The two largest groups are Group 2 and Group 1, which include 192 and 160 wards, respectively. Again, at ward level, it indicates that less deprived areas tend to be more cultural advantaged, and more deprived areas tend to be less cultural advantaged. We can also observe that the majority of wards in Group 2, which are more deprived and less advantaged in cultural spending, are located in East London. The relationship patterns between deprivation level, cultural expenditure, and network properties discovered in this preliminary analysis part lead to our hypotheses in the next section. And the four distinct groups of the {\tiny }neighbourhoods will be referred to in our following hypotheses evaluation. 

\begin{table}
	\centering
	\caption{Groups of London Wards in ANOVA Analyses}
	\label{table4}
	\begin{threeparttable}
		\scriptsize
		\normalsize
		\begin{tabular}{ccccc}
			\toprule
			& {\textbf{Group 1}} & {\textbf{Group 2}} & {\textbf{Group 3}} & {\textbf{Group 4}}
			\\
			\midrule
			Initial IMD & Less deprived & More deprived & More deprived & Less deprived \\
			\textit{CEA} & More advantaged & Less advantaged & More advantaged & Less advantaged \\
			Number & 160 & 192 & 88 & 114 \\
			\bottomrule
		\end{tabular}
	\end{threeparttable}
\end{table}

\begin{figure}[!h]
	\centering\includegraphics[width=4.5in]{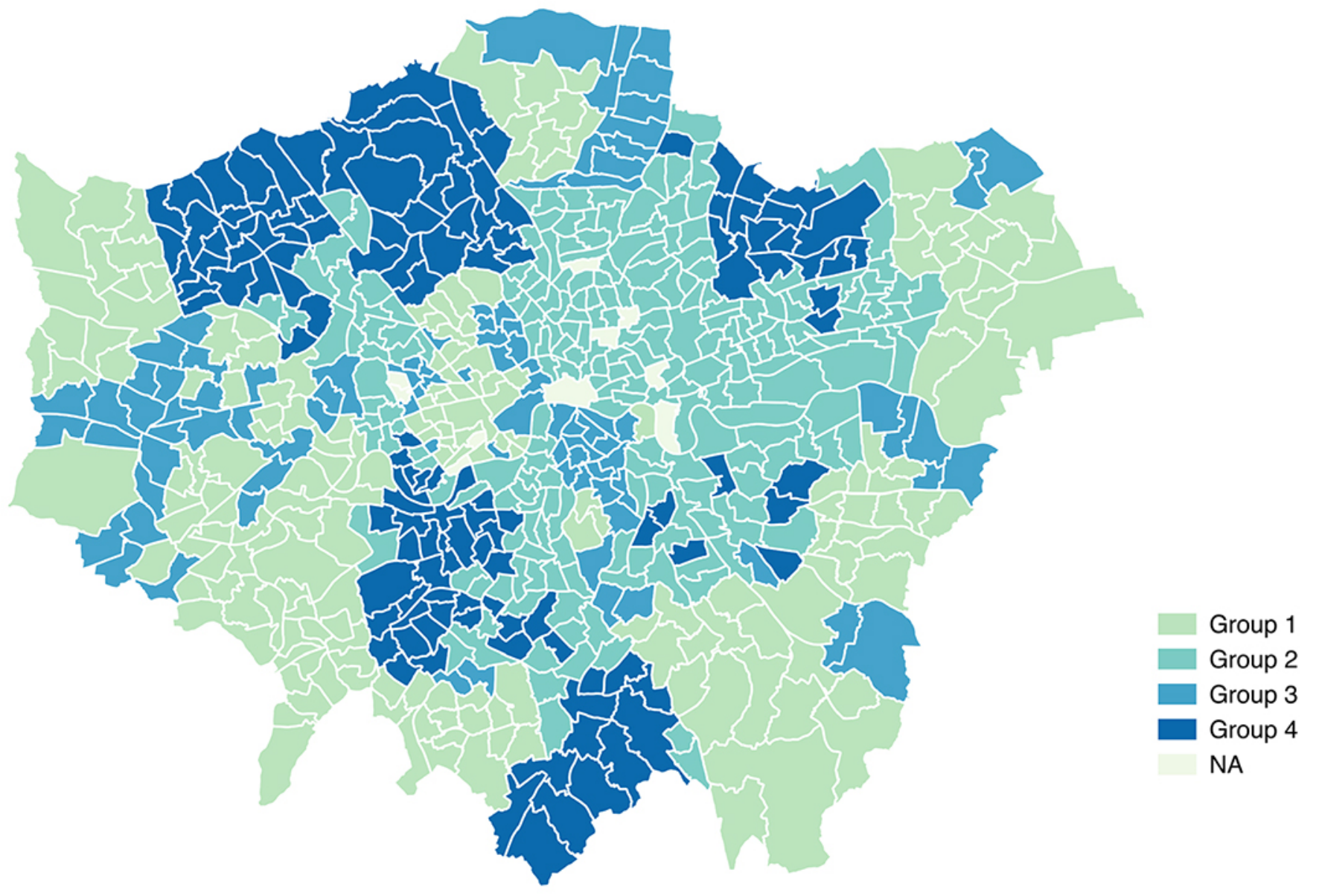}
	\caption{Spatial Distribution of Ward Groups in London}
	\label{fig5}
\end{figure}

\section{Hypotheses}
In order to explore the role that geo-social network data can play in culture-led urban regeneration study, we firstly reveal the underlying relationships between socio-economic status, local cultural expenditure, and geo-social network graph. As a base condition, we expect that different socio-economic conditions and amounts of cultural expenditure in neighbourhoods will lead to different network properties in the geo-social graph. Therefore, we hypothesise that: 

 \noindent \textbf{[H1]} Areas with high cultural investment and deprivation level \textit{have significantly different network and local properties} from areas with low cultural investment and deprivation level. 

This assertion lays the foundation of further investigation into the nature of culture-led urban regeneration, where based on existing case studies from literature\cite{1} and our preliminary analysis, we expect that cultural investment in more deprived areas results in growth. Specifically: 
   
 \noindent \textbf{[H2]} Areas with high cultural investment and deprivation level experience \textit{significant growth with respect to network and local properties} from areas with low cultural investment and deprivation level. 

Additionally, network studies have shown great potential in reflecting socio-economic conditions in communication networks and the prediction of deprivation\cite{3}. Based on this existing knowledge, we propose that network features such as centrality and clustering coefficient, combined with geographic and cultural expenditure factors are able to predict local socio-economic changes. We put forward the following two hypotheses to this end: 
    
 \noindent\textbf{[H3]} Network features of areas together with cultural expenditure and geographic features are powerful signals in predicting socio-economic change. 

 \noindent\textbf{[H4]} Network features of areas are better predictors of improvement than expenditure and geographic features.

\section{Evaluation \& Results}

In the following hypotheses evaluation, London's neighbourhoods will be studied at ward level and grouped according to Table 4 before two types of ANOVA analyses are run. We aim to get a deeper understanding of the relationship patterns we found at borough level, and test how they vary between ward groups and different time periods. We evaluate the last two hypotheses using a prediction framework, which allows us to reason about the predictive power of different features described in the metrics section. 

\subsection{[H1] Network and Local Properties}
To test the [H1], we employ independent one-way ANOVA to examine whether a statistically significant difference is found in terms of a set of network and local features between different ward groups. The value used for each feature is the average for the three years from 2011 to 2013. 

From the output results, we can see that there are significant effects of groups on six features at the $ p<.05 $ level. The feature that distinguishes less and more deprived groups is the average clustering coefficient with a statistically significant main effect of $ F(3,550)=4.15, p=.006 $. This shows that neighbourhoods from different socio-economic status groups presented significantly different clustering patterns in their network graphs. Furthermore, through taking a comparison between groups in Figure 6, we find that less deprived wards (Group 1 and Group 4) have higher means of average clustering coefficient than more deprived ones (Group 2 and Group 3), which illustrates that venues in less deprived areas are more likely to cluster together. Group 1, which represents less deprived and more cultural spending advantaged neighbourhoods, is the only group that exceeds the average of all wards (red dashed line in the figure).

\begin{figure}[!h]
	\centering\includegraphics[width=1.8in]{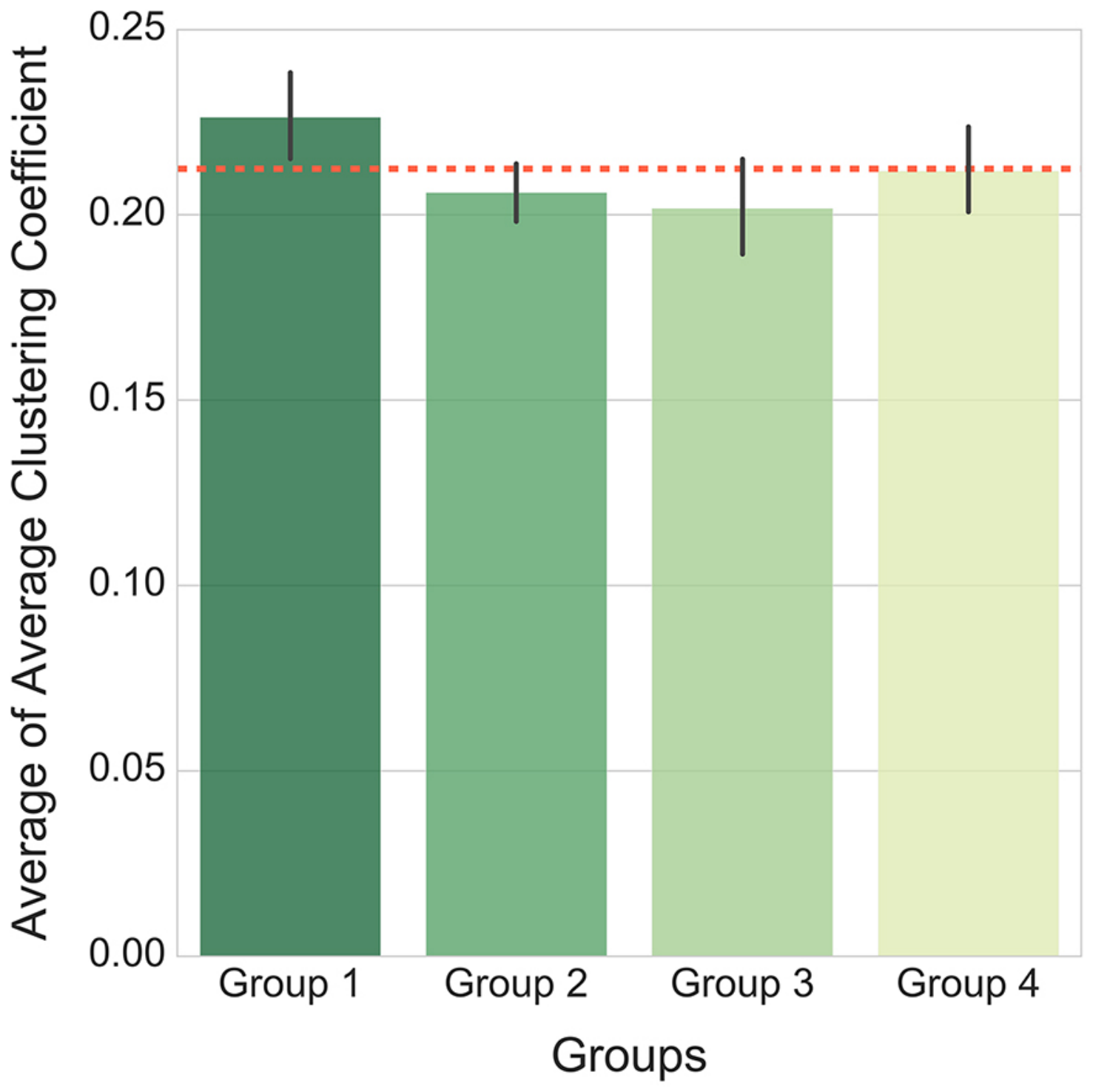}
	\caption{Means Plot for Average Clustering Coefficient in Independent One-way ANOVA Analysis}
	\label{fig6}
\end{figure}

The other five features that show statistically significant effects reveal differences between cultural advantaged and disadvantaged groups. The means of these five factors for different groups are plotted in Figure 7, from which we can find areas that gave a higher priority to culture (Group 1 and Group 3) had larger venue created number, node number, in-degree centrality, out-degree centrality, and venue created density on average. Moreover, Group 3, that was more deprived but invested relatively larger amount of money in culture from financial year 2010/11 to 2012/13, had the highest means in most cases. This result indicates that putting more effort into culture can lead to the stimulation of local business and the enhancement of vitality for urban areas. Additionally, a more striking effect can probably be seen in more deprived areas. 

\begin{figure}[!h]
	\centering\includegraphics[width=5.35in]{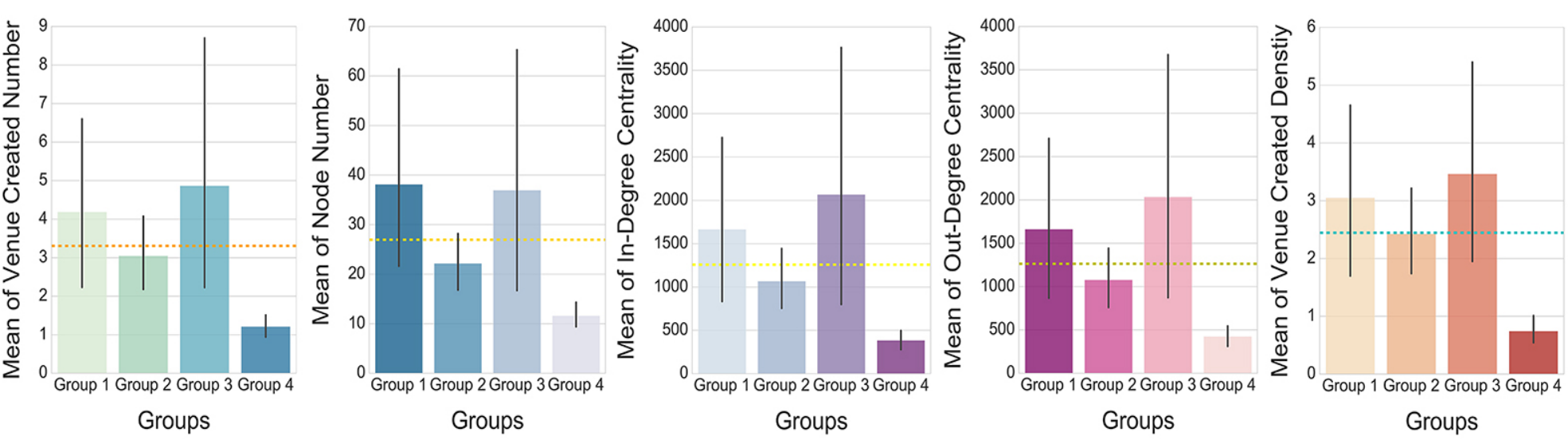}
	\caption{Means Plots for Variables with Statistically Significant Effects between Cultural Advantaged and Disadvantaged Groups in Independent One-way ANOVA Analysis}
	\label{fig7}
\end{figure}

Through One-way ANOVA analysis, we find that urban areas of different socio-economic status and cultural investment priorities vary in terms of local network features. Furthermore, areas with high cultural investment and deprivation level \textit{show significantly different values in network and local properties from those} areas with low cultural investment and deprivation level, which suggests that [H1] is true. 

\subsection{[H2] Growth of Network and Local Properties}
After discussing about the differences between groups on average, we test whether different groups of areas experienced significantly different growth patterns with respect to network and local properties in this subsection. Technically, we examine whether there are statistically significant differences between years and whether interaction effects exist between our two main factors, group and time, by factorial repeated measures ANOVA analysis. 

We present the means plots of five dependent variables that show statistically significant effects in Figure 8. From these plots, it can also be found that Group 1 and Group 3, which gave a high priority to culture had dramatic advantages in terms of almost all the features during the three years. Then, to consider the groups separately, Group 3 (more deprived and more advantaged in cultural investment) is still the one that had highest means in general, while Group 4 (less deprived and less advantaged) is the lowest with respect to all the features. These results demonstrate that significant differences not only exist between ward groups, but also show between different time points. It also reveals that the advantage of culture-supporting areas in various local and network properties is a dynamic and continuous process rather than an occasional phenomenon shown in a single year.
  
Moreover, significant interaction effects between group and year are found in three features: venue created density ($p=.008$), in-degree centrality ($p=.038$), and out-degree centrality ($p=.037$). We then run pairwise comparisons between different years and find that each pair of time points are observed to have statistical significance on venue created density, which reinforces that significant changes exist between years on this feature. As for the two centrality metrics, Group 1 differed significantly from Group 2 and Group 3, suggesting that less deprived neighbourhoods that spent more money on culture experienced significantly different changes with regard to in-flow and out-flow transitions compared to more deprived neighbourhoods.

\begin{figure}[!h]
	\centering\includegraphics[width=5.35in]{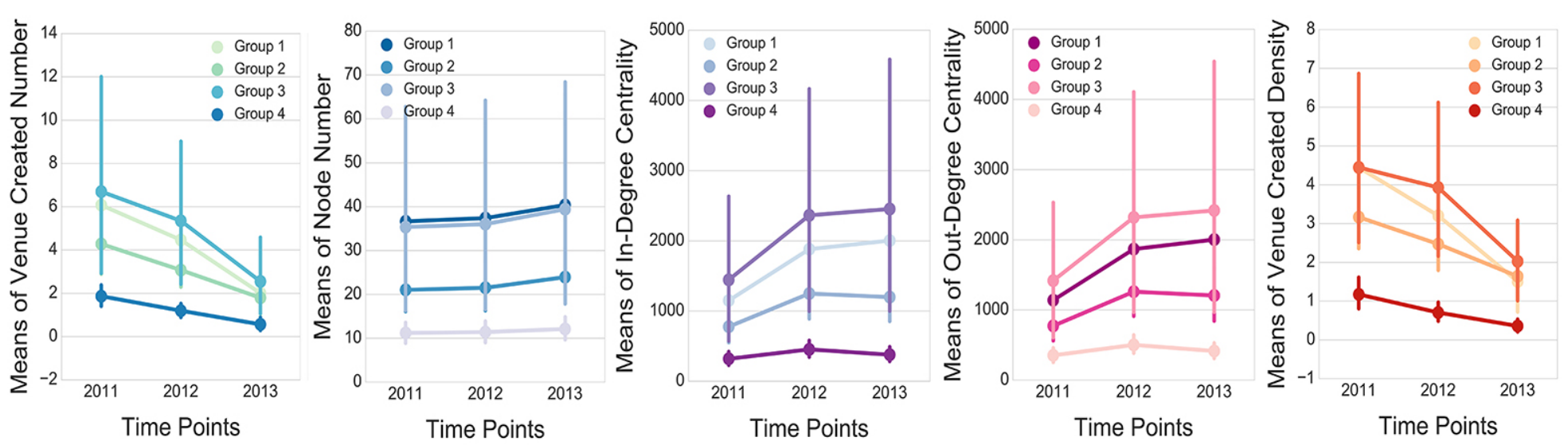}
	\caption{Means Plots for Variables with Statistically Significant Effects in Factorial Repeated Measures ANOVA Analysis}
	\label{fig8}
\end{figure}

With the help of factorial repeated measures ANOVA in this subsection, we detect how the growth of network and local properties varied between groups and time points. We confirm [H2] by finding that areas with high cultural investment and deprivation level experience significant growth with respect to network and local properties from areas with lower cultural investment and deprivation level. The interaction effect between group and time discovered can be further studied to explore whether there was a major culture-related policy or investment taking effect in certain group at certain time point.

In summary, the ANOVA analysis results for wards presented in the evaluation of [H1] and [H2] confirm the trend that we observed in the preliminary analysis for boroughs. Significant differences show in geo-social network variables between groups, time points, as well as their interaction effects, suggesting that urban areas with different socio-economic situations and cultural{\tiny } investment attitudes present different network graph patterns. Generally, investing more in culture is able to promote the growth of network graph in several ways. Additionally, this effect is more evident for those more deprived areas. 

Building on the findings from preliminary and ANOVA analyses, we introduce a supervised learning framework which exploits the prediction features displayed in Table 3 to predict IMD changes for London wards next. We assess whether our prediction features from three main categories of geographic, cultural expenditure, and network can be combined to build prediction models with good performance as a response to [H3]. Followed by the overall evaluation of classification models, our focus turns to explore the predictive power of different feature classes, especially the network feature set to test [H4]. 

\subsection{[H3] Prediction Model and Overall Evaluation}
In this subsection, we establish prediction models for the IMD change and discuss the performance of various methods on different neighbourhood sets. 

The target feature of our prediction is the binary IMD change in 2015 compared to the initial condition in 2010, which can either become more deprived or less deprived. We propose a supervised learning approach to tackle this binary classification problem: for each ward, we collect its initial IMD rank in 2010 and basic geographic features; We calculate the average local cultural expenditure of five kinds and the average \textit{CEA} during financial years from 2010/11 to 2012/13; Also, we compute local and network metrics for the beginning and end snapshots of 2011 and 2013, respectively, before calculating multiple growth rate features. The full list of prediction features we adopt to discriminate areas that are more or less likely became better or worse in terms of socio-economic condition has been presented in Table 3. After all these values are collected and calculated, we train classifiers and use a stratified 10-fold cross validation as the evaluation approach. The supervised learning methods we implement are classification tree, random forest, logistic regression, and Naive Bayes using the algorithms in library Scikit-learn\footnote{http://scikit-learn.org/stable/index.html}. After models are trained and established, we employ the AUC value, the average accuracy, and the average precision as measures to evaluate prediction performance for our classifiers. Precision is the fraction of positive predictions that are correct. Accuracy represents the proportion of the total number of predictions that are correct. AUC, which is not as intuitive as previous two measures, stands for area under ROC (receiver operating characteristic) curve and is commonly used as a measure of the overall quality of binary classification models.
Its value lies between 0.5 to 1, where a random classifier has a AUC of 0.5 and a perfect classifier's AUC is equal to 1\cite{22}. 

Instead of only focusing on the whole set of London's wards, we also class them according to how large their IMD rank changes are from 2010 to 2015, so as to discuss whether the prediction effectiveness varies when looking at areas with different IMD changes. In Figure 9, we present the IMD change distribution of wards in London. As we can see from the figure on the left, the IMD change of the whole ward set is normally distributed. Even though there are 625 London wards in total, 385 wards that have data available for all the prediction features are involved. The number of wards for each subset is also presented in Figure 9 on the right. Due to the consideration of sufficient sample size for our supervised learning models, besides the whole ward set, four subsets are also chosen to run the test, which are wards with the IMD rank change larger than 10, larger than 20, larger than 30, and larger than 40, respectively.

\begin{figure}[!tbp]
	\centering
	\subfloat{\includegraphics[width=0.49\textwidth]{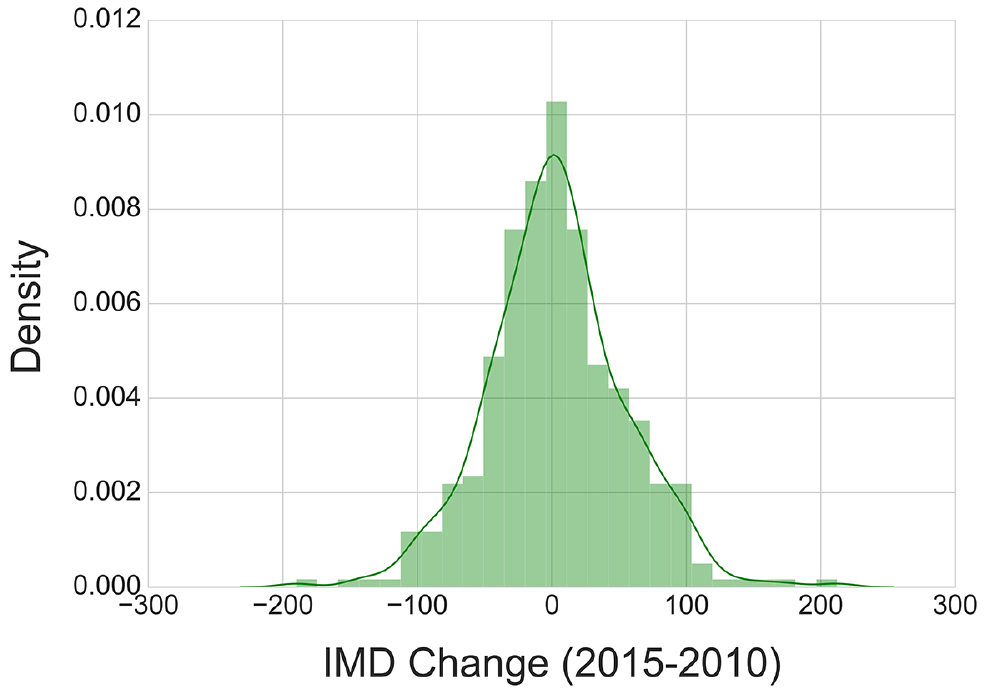}\label{fig9:f1}}
	\hfill
	\subfloat{\includegraphics[width=0.47\textwidth]{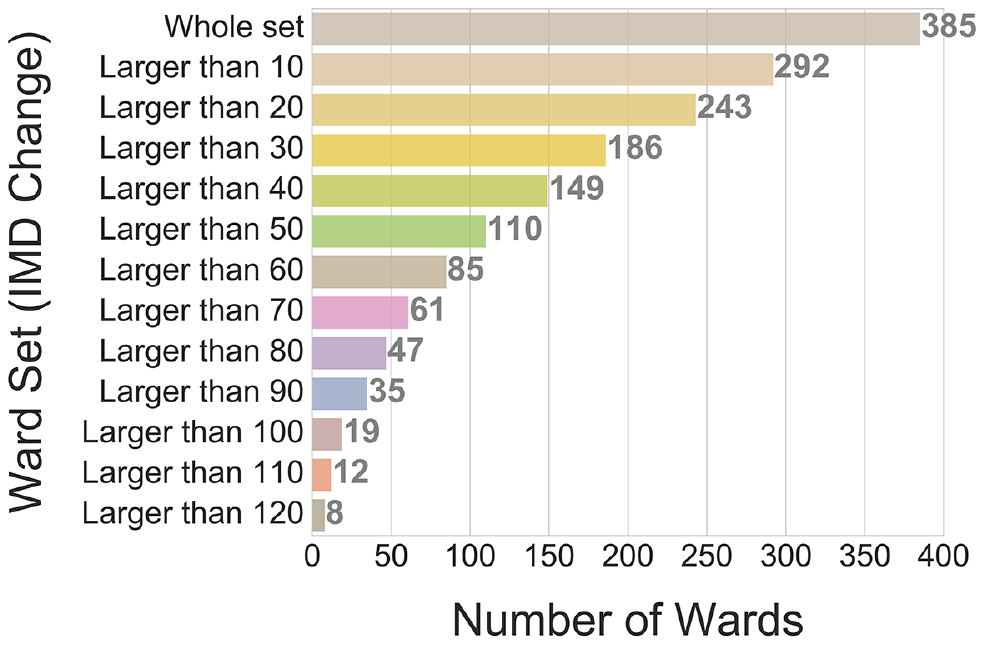}\label{fig9:f2}}
	\caption{Distribution of the IMD Change for London Wards}
\end{figure}

Overall, our prediction results shown in Figure 10 reveal that the inclusion of network, cultural expenditure, and geographic features offers high prediction performance by giving AUC scores over 0.7 for almost all the classifiers. And the best prediction performance shows when we look at wards with the IMD rank change larger than 40 using Naive Bayes, that the AUC reaches 0.865. As for the accuracy and precision measures, the scores are also higher than 0.7 in general. 

In addition, for different ward sets, we observe that it shows a rising tendency for all the classifiers in terms of evaluation measures in Figure 10. This finding suggests that better prediction results can be achieved from wards that have larger IMD changes. The reason for this is probably that neighbourhoods which experienced larger IMD changes showed more evident changes in local and network properties, making their socio-economic changes easier to be predicted.

When comparing the performance between different classifiers, we can see that Naive Bayes and random forest outperform the other two methods with high values in terms of all the three metrics. Followed by Naive Bayes and random forest, logistic regression performs slightly worse, whereas classification tree presents lowest values. While we have not explored exhaustively initialisation parameters of the four classifiers, what is important with regards to the goals of the present work, is that their performance evolves steadily with respect to the feature exploration we are demonstrating next. 

\begin{figure}[!h]
	\centering\includegraphics[width=5.35in]{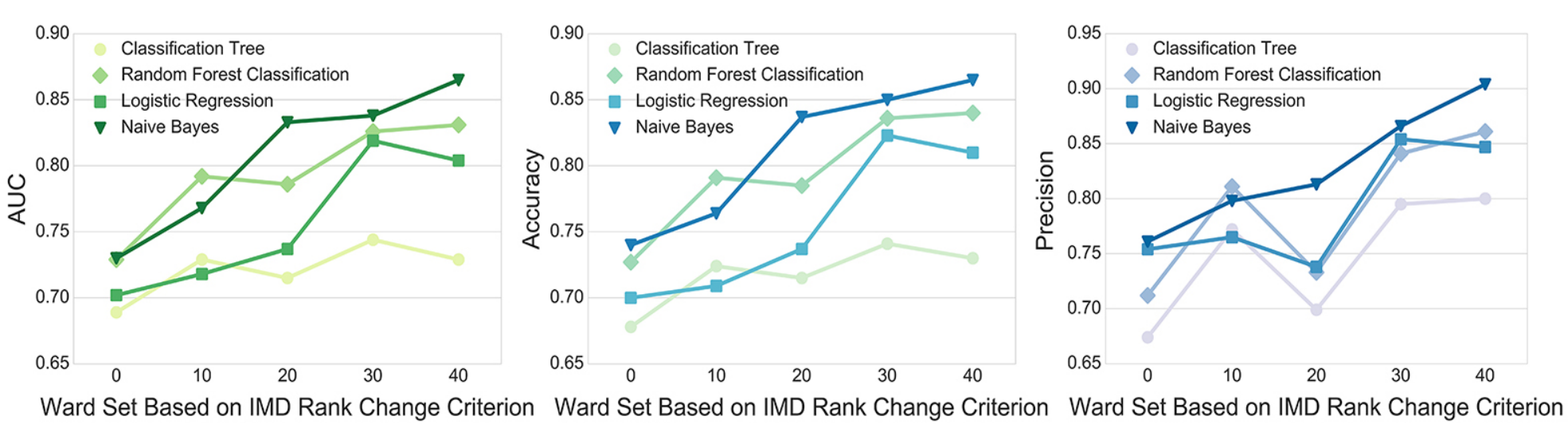}
	\caption{Evaluation for Supervised Prediction Methods on Different Ward Sets}
	\label{fig10}
\end{figure}

The high performance of our models presented by overall evaluation result shows the feasibility and superiority of utilising network features together with cultural expenditure and geographic features to predict the IMD change for urban areas [H3]. This desired outcome once again proves that crowdsourced data can play a significant role in socio-economic deprivation prediction instead of expensive census data.

\subsection{[H4] Individual Features Evaluation}

After evaluating the overall performance of prediction models, we investigate the predictive power of individual features in this part of analysis. 

Firstly, we study the predictive power of each individual feature in the prediction model. We take random forest classifier as an example and compute relative importance for each feature, which is determined in terms of the gini index \cite{23}. In Figure 11, we can see that \textit{CEA} (cultural expenditure advantage) plays the most significant role, being the only feature with an importance score over 0.1. It is followed by features \textit{GRIOR} (growth rate of ratio of in-degree centrality over out-degree centrality) and \textit{CEOP} (expenditure on open spaces per capita) with values around 0.09. While for the last two features, \textit{Sub region} (sub-region of London where a ward locates) and \textit{GRVC} (growth rate of venues created number), the importance scores are less than 0.03.

\begin{figure}[!h]
	\centering\includegraphics[width=5in]{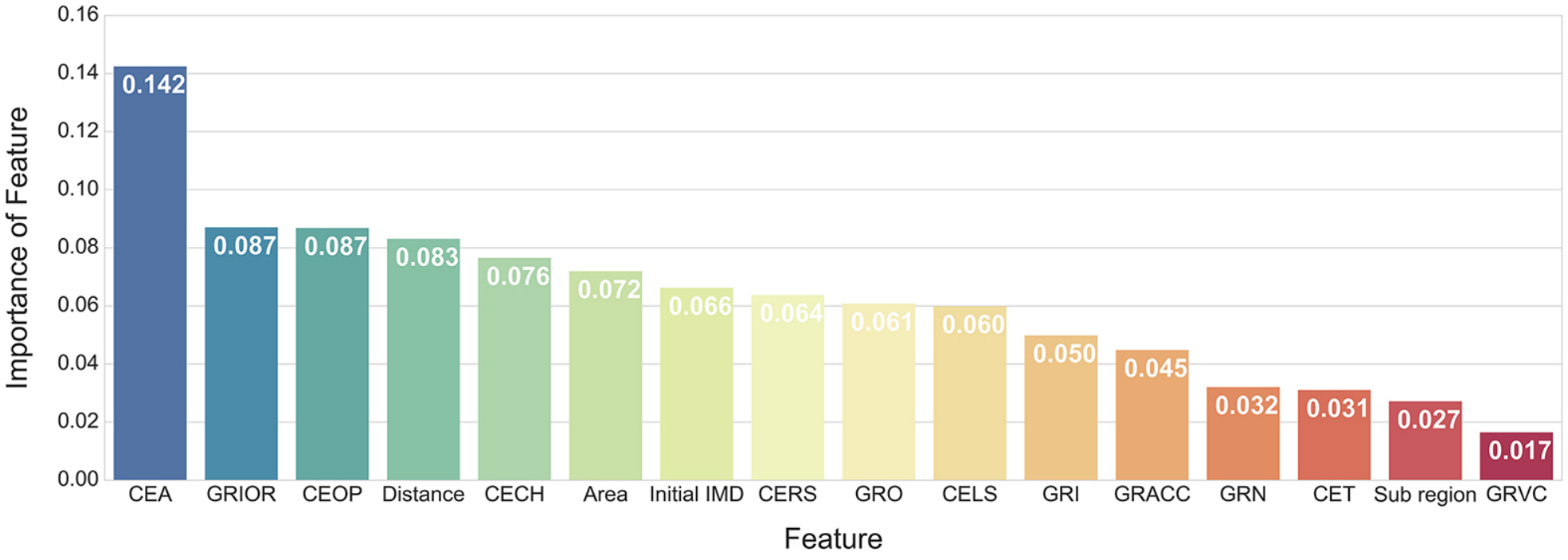}
	\caption{Relative Importance Evaluation for Each Feature in Random Forest Classification}
	\label{fig11}
\end{figure}

Next, in order to understand to which extent different feature classes are contributing to the prediction, we test what prediction performance can be achieved by removing one feature class with respect to the full model. The prediction results of these new models with two feature classes considered in each are shown in Figure 12. From the figure, we can see that geographic features make the smallest contribution to prediction models, as the reduction of prediction effectiveness is least when they are removed. In contrast, network features as a whole are the strongest signal to predict the IMD change [H4]. 

\begin{figure}[!h]
	\centering\includegraphics[width=5.35in]{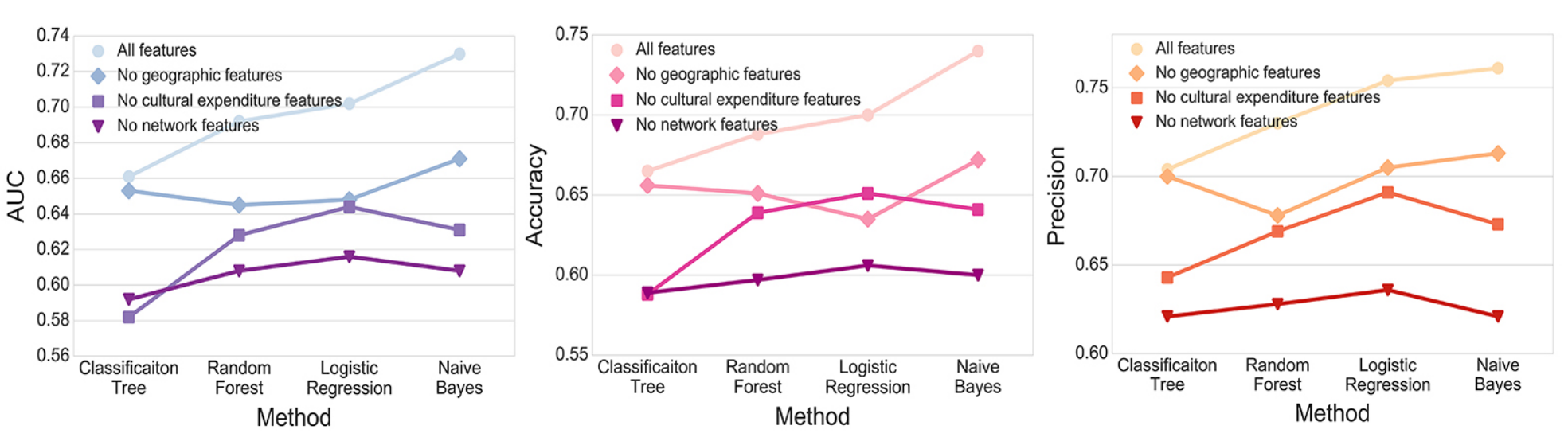}
	\caption{Evaluation for Supervised Prediction Methods on Different Feature Sets}
	\label{fig12}
\end{figure}

In conclusion, we present the feasibility and effectiveness of using geo-social network, cultural expenditure, and geographic features to infer whether the socio-economic status gets better or worth for small urban areas in the evaluation of [H3] and [H4]. The results of our prediction models are favourable with AUC scores higher than 0.7 in general. Moreover, the prediction performance sees an improvement when we focus on neighbourhoods that have larger IMD changes, and employ Naive Bayes and Random Forest classifiers. To evaluate the prediction contribution of features separately, network is the category that plays the most significant role in the prediction and cultural expenditure advantage is the most powerful individual feature.    
 
\section{Discussion \& Conclusions}

In this work, we have investigated the socio-economic impact of cultural expenditure on London neighbourhoods, visible through the lens of location-based mobile data. Finding evidence of the regenerative effects of such investment in local areas, we take a step further by trying to predict socio-economic impact based on geographic, network and cultural expenditure features. Overall, we have put forth evidence of the potential of using geo-social data for detecting and predicting the impact of culture-led regeneration strategies. This has a number of significant implications for location-based mobile systems, local governments and policy makers alike.

Firstly, we  have explored the relationship between socio-economic condition, cultural investment, and geo-social network graph, finding that spending more on culture can lead to an improvement of local development, especially for more deprived neighbourhoods. This observation verifies the effectiveness of implementing cultural strategies in urban regeneration project, and illustrates that culture-led regeneration policies are more suitable for underprivileged areas. On the basis of this finding, we suggest governments and policy makers taking socio-economic condition into consideration as an important factor when implementing cultural strategies to promote local development. Furthermore, our research has presented geo-social network data's ability to enrich or even replace traditional census-based deprivation statistics by proposing a supervised learning framework for predicting the outcome of cultural investment in London neighbourhoods. Although at present we perform our evaluation on annual snapshots, which nevertheless improves on current 5-year government census statistics, our future work will involve higher temporal resolution in order to explore these effects even further. Moreover, integrating predictive growth modelling in current urban planning systems could significantly help the government decide on the amount of cultural expenditure and along with geo-social network metrics, predict the impact of such investments.

One notable limitation of our research is that Foursquare data as well as other social media data in general is shown to be biased towards more central than peripheral parts of the city and often omits significant portions of the population who might be more deprived. Also, the Foursquare venues cannot represent the entire set of urban places exactly as how they present in the city, but they undeniably provide us an inspirational view to understand cities in fine grained spatio-temporal contexts. In our case, Foursquare data make it possible to  observe culture-led regeneration policies taking place and detect their effects. Furthermore, it has been shown to have interesting potential of uncovering gentrification processes in the city where more affluent residents tracked by Foursquare might replace the local deprived population \cite{13}. Additionally, cultural activities have been widely associated with benefits related to health, well-being and prosperity \cite{24}. One possible future application of our research is to the location-based application domain where recommendation systems can make use of geo-social and public data to help direct users to areas of 'cultural buzz' \cite{25}. In addition, although there are many other factors to take into account with regards to neighbourhood properties, our network and cultural advantage features provide the potential to also build a recommendation engine for residential neighbourhoods and where to buy property. As a whole, our work aims to shed light on the practicality of such future applications and invites further research into the exploration of culture-led urban development using digital traces.

\section*{Acknowledgment}

We would like to thank Foursquare for supporting this research by providing the dataset employed in the analysis.

\vspace{3ex}

\textbf{Ethics.} No special ethical permit or assessment was requested. Our paper does not involve animal ethics and fieldwork.

\textbf{Data Accessibility.} There are totally three data sources used in our research. In the dataset section, we describe these datasets and introduce their basic properties. Among them, the two official datasets can be download from the government's websites (https://www.gov.uk/government/statistics/english-indices-of-deprivation-2010, 

\noindent https://www.gov.uk/government/statistics/english-indices-of-deprivation-2015, 

\noindent https://www.gov.uk/government/collections/local-authority-revenue-expenditure-and-financing).
The Foursquare dataset has been shared by the company through an official agreement with the University of Cambridge that we have no authority to redistribute it.

\textbf{Competing Interests.} The authors declare no competing interests.

\textbf{Authors' Contributions.} X. Zhou, D. Hristova, A. Noulas, C. Mascolo designed the study together. X. Zhou carried out the data analysis, interpreted the results and wrote the manuscript; D. Hristova helped to draft and edit the paper; A. Noulas gave many suggestions, especially on the prediction part; C. Mascolo coordinated the study and helped to draft the manuscript; and M. Sklar helped with the Foursquare data. All authors gave final approval for publication.

\textbf{Funding.} Xiao Zhou is sponsored by the China Scholarship Council and the Cambridge Trust.


\end{document}